\documentclass[apj]{emulateapj}
\usepackage{graphicx}

\def\mathnew{\mathsurround=0pt}
\def\simov#1#2{\lower .5pt\vbox{\baselineskip0pt \lineskip-.5pt
       \ialign{$\mathnew#1\hfil##\hfil$\crcr#2\crcr\sim\crcr}}}
\def\simg{\mathrel{\mathpalette\simov >}}
\def\siml{\mathrel{\mathpalette\simov <}}
\def\gtrsim{\mathrel{\mathpalette\simov >}}
\def\lesssim{\mathrel{\mathpalette\simov <}}

\def\Mesz{M\'esz\'aros\,}

\def\msun{M_\odot}

\def\vareps{\varepsilon}

\def\eps{\epsilon}

\def\beq{\begin{equation}}
\def\enq{\end{equation}}
\def\bea{\begin{eqnarray}}
\def\ena{\end{eqnarray}}
\def\bec{\begin{center}}
\def\enc{\end{center}}

\def\blist{\begin{list}{$\bullet$}{\itemsep 0.0in \parsep 0.0in}}
\def\elist{\end{list}}
\def\bitem{\begin{list}{\arabic{enumi}.}{\usecounter{enumi} \itemsep 0.0in \parsep 0.0in}}
\def\eitem{\end{list}}
\def\cm{\hbox{~cm}}
\def\s{\hbox{~s}}
\def\erg{\hbox{~erg}}

\def\PeV{\hbox{PeV}}
\def\TeV{\hbox{~TeV}}
\def\GeV{\hbox{~GeV}}

\def\eV{\hbox{~eV}}

\def\part{\partial}

\def\pc{{\rm pc}}
\def\kpc{{\rm kpc}}
\def\Mpc{{\rm Mpc}}

\def\yr{{\rm yr}}

\def\m-pl{m_{Pl}}

\def\h75{h_{75}}
\def\Omh75{\Omega h^2_{75}}
\def\Omh70{\Omega h^2_{70}}

\def\G{{\rm G}}

\def\fun#1#2{\lower3.6pt\vbox{\baselineskip0pt\lineskip.9pt
  \ialign{$\mathsurround=0pt#1\hfil##\hfil$\crcr#2\crcr\sim\crcr}}}

\def\Ek525{E_{k,52.5}}
\def\varep{\vareps_p}

\newcommand{\Ecut}{$\epsilon_\gamma^{cut}$}
\newcommand{\Ebreak}{$\epsilon_\gamma^{br}$}

\begin{document}

\title{Extragalactic star-forming galaxies with hypernovae and supernovae\\ 
as high-energy neutrino and gamma-ray sources:\\ the case of the 10 TeV neutrino data}

\author{Nicholas Senno\altaffilmark{1}, Peter \Mesz\altaffilmark{1}, Kohta Murase\altaffilmark{1,2}, Philipp Baerwald\altaffilmark{1}, and Martin J. Rees\altaffilmark{3}}
\altaffiltext{1}{Dept. of Physics, Dept. of Astronomy \& Astrophysics, Center for Particle and
Gravitational Astrophysics, Pennsylvania State University, University Park, PA 16802, USA}
\altaffiltext{2}{Institute for Advanced Study, Princeton, NJ 08540, USA}
\altaffiltext{3}{Institute of Astronomy, University of Cambridge, Cambridge CB3 0HA, U.K.}

\email{nbs5044@psu.edu}




\begin{abstract}

In light of the latest IceCube data, we discuss the implications of the cosmic ray (CR) energy input from 
hypernovae and supernovae into the Universe, and their propagation in the hosting galaxies and galaxy 
clusters or groups.  The magnetic confinement of CRs in these environments may lead to efficient neutrino production via $pp$ 
collisions, resulting in a diffuse neutrino spectrum extending from PeV down to 10 TeV energies, 
with a spectrum and flux level compatible with that recently reported by IceCube. 
If the diffuse 10 TeV neutrino background largely comes from such CR reservoirs, 
the corresponding diffuse $\gamma$-ray background should be compatible with the recent \textit{Fermi} data. 
In this scenario, the CR energy input from hypernovae should be dominant over that of supernovae, 
implying that the starburst scenario does not work if the supernova energy budget is a factor of 
two larger than the hypernova energy budget.
Thus, this strong case scenario can be supported or ruled out in the near future. 

\end{abstract}

\maketitle

\section{Introduction}
\label{sec:intro}

The detection of PeV and sub-PeV astrophysical neutrinos by IceCube \citep{2013PhRvL.111b1103A,IC3+13pevnu2,2014PhRvL.113j1101A},
which more recently has been extended down to 10 TeV energies \citep{IC3+15tevnu}, is a major
development. The origin of these neutrinos is a matter of intense interest \citep[for recent
reviews, see e.g.,][]{2014NuPhS.256..241M,Murase14pevrev}. 
Star-forming galaxies, especially starbursts, are promising candidates
\citep{Loeb+06nustarburst,Murase+13pev,2013arXiv1311.0287K,Tamborra+14igbnu,Chang+14pevnugam,Anchordoqui+14pevnu}, 
in which the major contributing sources may be supernovae (SNe), as well as their hyper-energetic equivalent the 
so-called hypernovae (HNe) \citep{2003A&A...409..799S,Fox+13pev,He+13pevnuhn,Murase+13pev,Liu+13pevnuhn,Ahlers+14pevnugal} 
or active galactic nuclei (AGN) \citep{Murase+14nuagn,Tamborra+14igbnu}.
Other possible candidates include low-luminosity classes of $\gamma$-ray bursts \citep{2013PhRvL.111l1102M,2013ApJ...766...73L}, 
radio-loud active galactic nuclei
\citep{1991PhRvL..66.2697S,2004PhRvD..70l3001A,Murase+14nuagn,2014JHEAp...3...29D,2014arXiv1410.2600K}, galaxy clusters and groups \citep{Murase+08clusternu,Murase+13pev} 
with accretion shocks \citep{Keshet+04igshock,Inoue+05igshock} that may 
accelerate cosmic rays (CRs) to higher energies 
\citep{Hillas84rev,Biermann+87agncr} or other CR sources such as galaxy mergers 
in clusters \citep{Kashiyama+14pevmerg}.  

In this work, we will concentrate mainly on the HN/SN origin of neutrinos from $pp$ 
interactions in the starburst and normal star-forming intragalactic material and the intracluster medium.  
In particular, we discuss their implications in light of the latest IceCube data in the 10 TeV range. Constraints from the diffuse $\gamma$-ray background measured by {\it Fermi} are even more pronounced in 
this case.

\section{Hypernova and supernova energy input rate}
\label{sec:hn}

\setcounter{footnote}{1}

HNe typically belong to a sub-class of broad-line Type Ib/c SNe with ejecta kinetic energies 
of order $E_{k}=10^{52}E_{k,52}\erg$, representing a fraction $\xi_{hn} \simeq 
4\times 10^{-2}\, \xi_{hn,-1.4}$ of all core-collapse SNe, with substantial uncertainties 
\citep{2010ApJ...721..777A,Guetta+07grbhnrate,2011MNRAS.412.1522S}.  The rate of all core-collapse SNe 
is $1.06\times 10^{-4}\, \Mpc^{-3}\yr^{-1}$ \citep[e.g.,][]{Taylor+14ccsnrate}, which implies 
a local HN rate of ${\cal{R}}_{hn}\sim4\times 10^{-6} \xi_{hn,-1.4}\Mpc^{-3}\yr^{-1}$.
If the fraction of HN remnant kinetic energy transferred to CRs is $E_{cr,hn} \simeq 2.8\times10^{51}\;\mathrm{ergs}$, the CR energy density input rate  in the Universe is 
$\dot{U}_{cr}\simeq E_{cr,hn} {\cal{R}}_{hn}  
\simeq 1.2\times 10^{46}\, \xi_{hn,-1.4}\,E_{cr,51.4} \,\erg~\Mpc^{-3}\yr^{-1}$. 
Furthermore, if the CRs are protons with a power law 
distribution $N(\vareps_{p})\propto \vareps_p^{-\Gamma}$ between $\vareps_{p,min} \sim 1\GeV$ 
and $\vareps_{p,max}\siml 10^{17}\vareps_{p,17}\eV$, the energy density input rate per logarithmic 
interval of energy is $\vareps_p Q_{\vareps_p} \simeq {\dot U}_{cr}/{\cal C}~\erg~\Mpc^{-3}\yr^{-1}$, 
where the bolometric correction is ${\cal C}= \ln{(\vareps_{p,max}/ \vareps_{p,min})}$ for a 
spectral index $\Gamma=2$\footnote{It would be ${\cal C}= \left[1-(\vareps_{p,max}/
\vareps_{p,min})^{2-\Gamma}\right] \vareps_{p,min}^{2-\Gamma} (\Gamma-2)^{-1}$ if $\Gamma\neq 2$.}. 
Assuming $\Gamma\sim 2$ and taking ${\cal C}\sim 18 {\cal C}_{18}$ the local ($z=0$) CR 
energy input per logarithmic interval in the Universe due to hypernovae is
\bea
\left(\vareps_p Q_{\varep}\right)_{hn} \simeq 6.4 \times 10^{44}\, \xi_{hn,-1.4} \,{\cal C}_{18}^{-1} E_{cr,hn,51.4} \cr  \erg~\Mpc^{-3}\yr^{-1}, ~~~~~~~~~
\label{eq:crinput}
\ena
which is larger than the typical value expected for $\gamma$-ray bursts.
Conventional SNe will also contribute significantly to lower-energy CRs, 
having a smaller kinetic energy input $E_{k,sn}\simeq 10^{51}E_{ksn,51}\erg$ but a larger
rate ${\cal R}_{sn}$. The typical CR energy of SNe is uncertain and could be less (e.g. $E_{cr,sn} = 4.8\times10^{49}\;\mathrm{erg}$). In general, the energy injection rate for SNe is given by
\beq 
\label{eq:crinputsn}
\left(\vareps_p Q_{\varep}\right)_{sn} = \frac{(1-\xi_{hn})}{\xi_{hn}}\,
\frac{\mathcal{C}_{hn}}{\mathcal{C}_{sn}}\,\frac{E_{cr,sn}}{E_{cr,hn}}
\left(\vareps_p Q_{\varep}\right)_{hn}\, .
\enq
It is believed that SNe can typically accelerate CRs to a maximum energy $\vareps_{p,max}\sim 
10^{15}\;\mathrm{eV}$ resulting in $\mathcal{C}_{sn} \sim 13.8$. The energy input due to conventional 
SNe would be typically larger than that of HNe at lower energies.  However, with the parameters given above, 
Eq. (\ref{eq:crinputsn}) implies the energy input rate of SNe for CRs with $\vareps_p \lesssim \vareps_{p,max,sn}$ is roughly half that of HNe. Below, we leave this ratio as a free parameter.

\section{Shock acceleration}
\label{sec:accel}

A typical Type Ibc SN has a bulk ejecta mass of $M_{ej}\simeq3 M_{ej,0.5}\msun$, and a HN has an average velocity $\beta_{ej}= (V_{ej}/c)= 6.1\times 10^{-2}E_{k,hn,52}^{1/2}M_{ej,0.5}^{-1/2}$.    
The postshock random magnetic field strength is expected to be amplified to a fraction 
$\eps_B$ of the postshock thermal energy, $B_s \sim(16\pi \eps_B n_g m_N c^2 \beta_{ej}^2)^{1/2}$,
 with $n_g$ being the interstellar particle number density. The upstream magnetic field should also be amplified by e.g., CR-streaming instabilities, but in 
any case the stronger magnetic fields in starburst galaxies may also be enough~\citep{Murase+13pev}.
Diffusive shock acceleration in the blast wave leads to a power law spectrum distribution
$N(\varep) \propto \varep^{-\Gamma}$, typically with $\Gamma\simg 2$, up to a maximum 
energy $\vareps_{p,max} \simeq (3/20) Z e B_s R_{dec} \beta_{ej}$ \citep{1983RPPh...46..973D},  or
\bea
\vareps_{p,hn,max}\simeq 10^{17} Z n_{g,2.3}^{1/6} 
   E_{k,hn,52}\, M_{ej,0.5}^{-2/3} 		\eV
\label{eq:epmaxdec}
\ena
for CRs with charge $Z$. However, in the following work we only consider CR protons. The above equation implies that the CRs are accelerated to $\sim 100\,\PeV$ as the shock slows and enters the so-called Sedov-Taylor phase. The maximum CR energy is expected to decrease with time during this deceleration phase \citep{drury2011escaping}. Also note that while many SNe and HNe may happen in relatively low-density regions such as superbubbles, the dependence on $n_g$ is weak.
For normal SNe, using the same parameters except for $E_k=10^{51}\erg$, the maximum
energy would be $\vareps_{p,sn,max}\simeq 1.1\times 10^{16} Z n_{g,2.3}^{1/6}
E_{k,51} M_{ej,0.5}^{-2/3}~~\eV$. 
As we will show below, HNe that occur in starburst galaxies can accelerate the majority of CRs 
which produce detectable high-energy neutrinos. 
Although we use typical numbers for HNe for our estimates, in star-forming galaxies hosting an AGN
the latter may be also contribute as a 10-100 PeV CR accelerator 
\citep{Murase+14nuagn,Tamborra+14igbnu}.

CRs suffer energy losses both during acceleration and after escaping their source.
Synchrotron losses are negligible at the energies considered here, the dominant loss
mechanism being hadronuclear ($pp$) collisions. 
The effective optical depth to $pp$ collisions undergone 
while advected downstream of the blast wave is $\tau_{pp,s} \sim t_{dyn}/t_{pp}\sim \kappa \sigma_{pp} R (c/V)$,
where $\kappa\sim 0.5$ is the inelasticity and $\sigma_{pp}({\mathbf \vareps_{p} = 100\;\mathrm{PeV}})\sim 10^{-25}\cm^2$ \citep[in the numerical calculations presented below, we use the energy dependent inelastic $pp$ cross section presented in][]{Kelner+06ppint}. Thus
$\tau_{pp,s} \sim 1.3\times10^{-6}\,E_{k,hn,52}^{-1/2}\,M_{ej,0.5}^{5/6}\,n_{g,2.3}^{-1/3}$, which is negligible compared to losses
during the subsequent propagation. Similar considerations apply also to the supernovae.

\section{Propagation effects and $pp$ optical depth}
\label{sec:prop}

The propagation of the CRs in the turbulent magnetic field of the host galaxy 
and galaxy cluster depends, in the diffusion approximation, on the strength of the
magnetic field $B$, the CR Larmor radius $r_L$, and the coherence length $\ell_c$ of the magnetic
field fluctuations. At the highest energies $\varep$, where $r_L(\varep)\gg \ell_c$, 
the CR diffusion coefficient is $ D(\varep) \propto r_L(\varep)^2 $. At lower energies, where 
$r_L(\varep)\siml \ell_c$, the diffusion coefficient is $D(\varep)\propto r_L(\varep)^{\alpha}$, 
where $\alpha=1/3$ ($1/2$) for a Komolgoroff (Kraichnan) fluctuation power spectrum \citep[e.g.,][]{Berezinsky+97clucr}. The two regimes can be interpolated as
\beq
D(\varep)=D_\ast \left[ \left({\varep}/{\vareps_{p,\ast}}\right)^{\alpha} 
  + \left({\varep}/{\vareps_{p,\ast}}\right)^{2} \right]
\label{eq:difcoef}
\enq
where $r_L(\vareps_{p\ast})=\ell_c/5$ with $D_\ast \simeq (1/4) c r_L(\vareps_{p\ast})$ 
\citep{Parizot04gzk}. Below we shall use $\alpha=1/3$ as an example, but the discussion
can be generalized to a general positive $\alpha$ value.

After leaving the source (e.g. HNe or SNe), the CRs first propagate diffusively through the 
host galaxy or are advected away by a strong galactic wind, with typical velocities of $V_w \sim 
1500\;\mathrm{km/s}$ in starburst galaxies \citep{Strickland+09M82wind} and $V_w \sim 500 
\;\mathrm{km/s}$ for normal star-forming galaxies \citep{Crocker12mwwind,Keeny+06mwwind}.
For a starburst galaxy (SBG) the gas scale height $H_g \sim 30-300\,\pc$ may be 
parameterized as $H_{sbg}\sim300\, pc \simeq 10^{21}H_{21}\cm$.  We assume a magnetic 
field strength of $B_g \sim 200\times10^{-6}B_{g,-3.7}\,\G$
and a coherence length parameterized here as $\ell_{c,g}\sim 10^{-1}H_g \sim 30 \pc 
\simeq 10^{20}\ell_{g,20}\cm$.  
Both quantities are subject to large uncertainties and
variations, so that the diffusion coefficient adopted here corresponds to the optimistic case.
For our fiducial starburst galaxy parameters, we obtain $\vareps_{p\ast,g} \sim 1.11\times 
10^{18}\ell_{g,20}B_{g,-3.7}\eV$ and $D_{\ast,g}\sim1.4\times 10^{29}\ell_{g,20}\cm^2/\s$. To ensure CR confinement, 
we require the coherence length to satisfy $\vareps_{p\ast,g} \gtrsim10-100$~PeV 
\citep{Murase+13pev}.
For a normal star-forming galaxy (SFG), we take the typical scale height to be $H_g \sim 1000\,\pc$, 
with $\ell_{c,g}\sim 10^{-1}H_g$, a magnetic field of $B_g \sim 6\mu$G \citep{2008AIPC.1085...83B} 
and interstellar medium density of $n_g \sim 1\;\mathrm{cm^{-3}}$.

Given the above, the time for CR diffusive escape from the galaxy can be calculated, which for starburst galaxies is  $t_{d,g}=H_g^2/6D_g \simeq 1.5 \times 10^{12} H_{g,21}^2 \ell_{g,20} B_{g,-3.7}^2 
\vareps_{p,17.2}^{-1/3}\s$.  The time for advective escape is $t_{w,g}  = 
H_g/V_w \simeq 6.2\times10^{12}\,H_{g,21}\,V^{-1}_{w,3.2}\;\mathrm{s}$ regardless of the CR energy. 
Notice that advective escape dominates diffusive escape from the galaxy for CRs with energy 
$\vareps_p \lesssim \vareps_w$ with 
\begin{equation}
\label{eq:epsw}
\vareps_w = \frac{Z\,e\,B\,l_c^{1-1/\alpha}}{5}\,\left(\frac{10\,V_w\,H_g}{3\,c}\right)^{1/\alpha},
\end{equation}
yielding $\vareps_w \sim 5.1\times10^{15}\;\mathrm{eV}$ for the fiducial parameters used here. 

The effective $pp$ optical depth undergone during propagation in a starburst galaxy is
$\tau_{pp,g}\simeq n_g \kappa \sigma_{pp} c \,\min [t_{d,g},t_{w,g}]$ or
\begin{eqnarray}
\label{eq:tauppg}
\tau_{pp,g} &\sim &  4.9\times10^{-3} \;n_{g,2.3}H_{g,21}^2\ell_{g,20}B_{g,-3.7}^2 \vareps_{p,19}^{-2}
\nonumber \\
\tau_{pp,g} &\sim &  0.55 \quad n_{g,2.3} H_{g,21}^2 \ell_{g,20} B_{g,-3.7}^2 \vareps_{p,17.2}^{-1/3} 
\nonumber\\
\tau_{pp,g} &\sim &  1 \quad n_{g,2.3} H_{g,21} V_{w,3.2} 
\end{eqnarray}
in the ranges $(\varep >\vareps_{p\ast,g})$, $(\vareps_w <\varep <\vareps_{p\ast,g})$ and
$(\vareps_p < \vareps_w)$, respectively. 

One can see that starburst galaxies are efficient neutrino factories via $pp$ interactions due to 
their high interstellar gas density. As seen in Fig. 1-4, normal star-forming galaxies have lower values of $\vareps_{p\ast,g}$, $\vareps_w$, and $\tau_{pp,g}$ resulting in only a modest amount of neutrinos produced at high energies. We will show below that if the starburst fraction is 
high and CRs with energies up to $\sim10-100$~PeV are sufficiently confined, the majority of the observed 
high-energy diffuse neutrino flux can be explained using HNe or other sources in 
starburst galaxies.

For the subsequent propagation in the galaxy cluster or group, the average magnetic field and coherence 
length are parameterized as $B_{cl}\sim 10^{-6}B_{cl,-6}\G$ and $\ell_{c,cl}\sim 30\kpc =
10^{23}\ell_{23}\cm$. 
This implies $\vareps_{p\ast,cl}\sim5.6\times 10^{18} Z\ell_{23}B_{cl,-6}\eV$ and
$D_{\ast,cl}\sim1.4\times 10^{32}Z\ell_{23}\cm^2\s^{-1}$. For a  cluster of $10^{15}\msun$
the virial radius is $R_{cl}\sim2.6M_{15}^{1/3}\Mpc \simeq 8\times 10^{24}M_{15}^{1/3}\cm$ 
and the diffusion time is $t_{d,cl}=R_{cl}^2/6D$.
At the maximum proton energy
this is $t_{d,cl}(\vareps_{p,max})\sim 2.3\times 10^{17}M_{15}^{1/3}\ell_{23}^{-2/3}
B_{cl,-6}^{1/3}Z^{1/3}n_{cl,-4}^{-1/18} \eps_{B,-2}^{-1/6}\Ek525^{-1/3} M_{ej,1}^{2/9} \\
\vareps_{p,16.94}^{-1/3}~\s$.
Similarly to the galactic component mentioned above, there is a spectral break when the diffusion time 
exceeds the injection time of CRs \citep{Murase+08clusternu}. 
Assuming CR injection effectively occurs during the Hubble time at the corresponding redshift  (i.e. $t_{age}(z) = \int_z^\infty\,dz'\left|\frac{dt_\ast}{dz'}\right|$), for a cluster located at redshift $z=1$ such a break occurs at an energy $\vareps_{p,cl}\sim4\times10^{17}Z\eV$.
If the cluster break energy is higher than the maximum HNe energy, CR diffusion does not significantly 
affect the fraction of CRs that interact in the intracluster medium. The cluster $pp$ optical depth 
is again $\tau_{pp,cl}=n_{cl} \kappa\sigma_{pp}c \,\min[t_{d,cl},t_{age}]$, assuming a typical 
intracluster gas density $n_{cl}\simeq 10^{-4}n_{cl,-4}\cm^{-3}$, at high redshifts (e.g. $z = 1$) 
$t_{age} \lesssim t_{d,cl}(\vareps_{p,max})$ so that
\beq
\tau_{pp,cl}(\vareps_{p,max})\sim 2.7\times 10^{-2} n_{cl,-4} 
(t_{age}/5.8~{\rm Gyr})
\label{eq:tauclumax}
\enq
For more nearby clusters $\tau_{pp,cl}$ increases as the cluster age approaches the local Hubble time, 
although the density is also redshift-dependent.
%

For a diffusion exponent $\alpha$ different from the value 1/3 used as an example above, the 
values of $\tau_{pp,g}$, $\tau_{pp,cl,H}$, etc., are calculated similarly and are somewhat 
different, as can be seen in the numerical results discussed in the next  section.

Further $pp$ collisions occur in the intergalactic medium after the CRs escape the cluster,
but with the intergalactic target density $n_{igm}=2.5\times 10^{-7}(\Omega_b h^2/0.022)\;\mathrm{cm^{-3}}$,
and a total flight time limited by $t_H\sim 10^{10}\yr$, the corresponding $\tau_{pp,igm}$
is negligible compared to the previous two contributions.

\section{Diffuse neutrino flux}
\label{sec:difnu}

When high-energy CRs undergo $pp$ interactions with the ambient intragalactic and intracluster medium, 
charged and neutral pions are created which subsequently decay to neutrinos and  $\gamma$-rays 
respectively. On average, the resulting neutrino and parent CR energies can be related by 
$\vareps_\nu \sim 0.03-0.05\,\vareps_p$. As a result, the diffuse neutrino flux (per flavor per 
logarithmic interval of energy) can be estimated using the CR energy injection
rate similarly to what is done for GRBs \citep{Waxman+97grbnu,Murase+13pev}, as
\beq
\vareps_\nu^2 \Phi_{\vareps_\nu} = \frac{c}{4\pi} \int_0^z \sum_i\frac{f_{i,pp}}{6}
\frac{\left(\vareps_p Q_{\vareps_p}\right)_{phys}}{(1+z')^4}\left|\frac{dt}{dz'}\right| dz',
\label{eq:fluxnugen}
\enq
where the physical CR energy injection rate per energy bandwidth at a given redshift $z$ is related 
to Eq.(\ref{eq:crinput}, \ref{eq:crinputsn}), cosmological evolution is taken into account by the 
scale factor $S(z)$ so the normalized physical star formation rate is
\bea
& \left(\vareps_p\,Q_{\vareps_p}\right)_{phys}(z) =  ~~~~~~~~~~~~~~~~~~~~~~~~~~~~~~~~~~~~\cr
&~~~\left[\left(\vareps_p Q_{\varep}\right)_{hn} + \left(\vareps_p Q_{\varep}\right)_{sn}\right]\,(1+z)^3\,S(z)
\ena
with 
\beq
S(z)=\left[(1+z)^{a\eta} + \left(\frac{1+z}{B}\right)^{b\eta} 
+ \left(\frac{1+z}{C}\right)^{c\eta}\right]^{1/\eta},
\label{eq:sfr}
\enq
where $a = 3.4,\, b = -0.3,\, c = -3.5, \, \eta \approx -10, \, B \simeq 5000 ,\, \mathrm{and}\, 
C \simeq 9$ \citep{2006ApJ...651..142H,Yuksel+08sfrgrb}.

The sum in Eq. (\ref{eq:fluxnugen}) is over the different galactic and cluster/group contributions. We assume a fraction $\xi_{sbg}$ 
of neutrinos are produced in starburst galaxies with the rest $\xi_{sfg} = 1-\xi_{sbg}$ produced 
in normal starforming galaxies: 
\begin{eqnarray}
\label{eq:fpp}
f_{pp,sbg} &=& \xi_{sbg}\left(1-e^{-\tau_{pp,g,sbg}}\right)\nonumber\\
f_{pp,sfg} &=& \xi_{sfg}\left(1-e^{-\tau_{pp,g,sfg}}\right)
\end{eqnarray}
\begin{eqnarray}
f_{pp,cl} = \left(1-e^{-\tau_{pp,cl}}\right)\,\, \times ~~~~~~~~~~~~~~~~~~~~\nonumber\\
\left[\xi_{sbg}\,e^{-\tau_{pp,g,sbg}} + \xi_{sfg}\,e^{-\tau_{pp,g,sfg}}\right]\nonumber
\end{eqnarray}
Note that in the last line of Eq. (\ref{eq:fpp}) only CRs which escape from the galaxies can contribute to the cluster component. 

\begin{figure}[t]
\centerline{\includegraphics[width=0.45\textwidth]{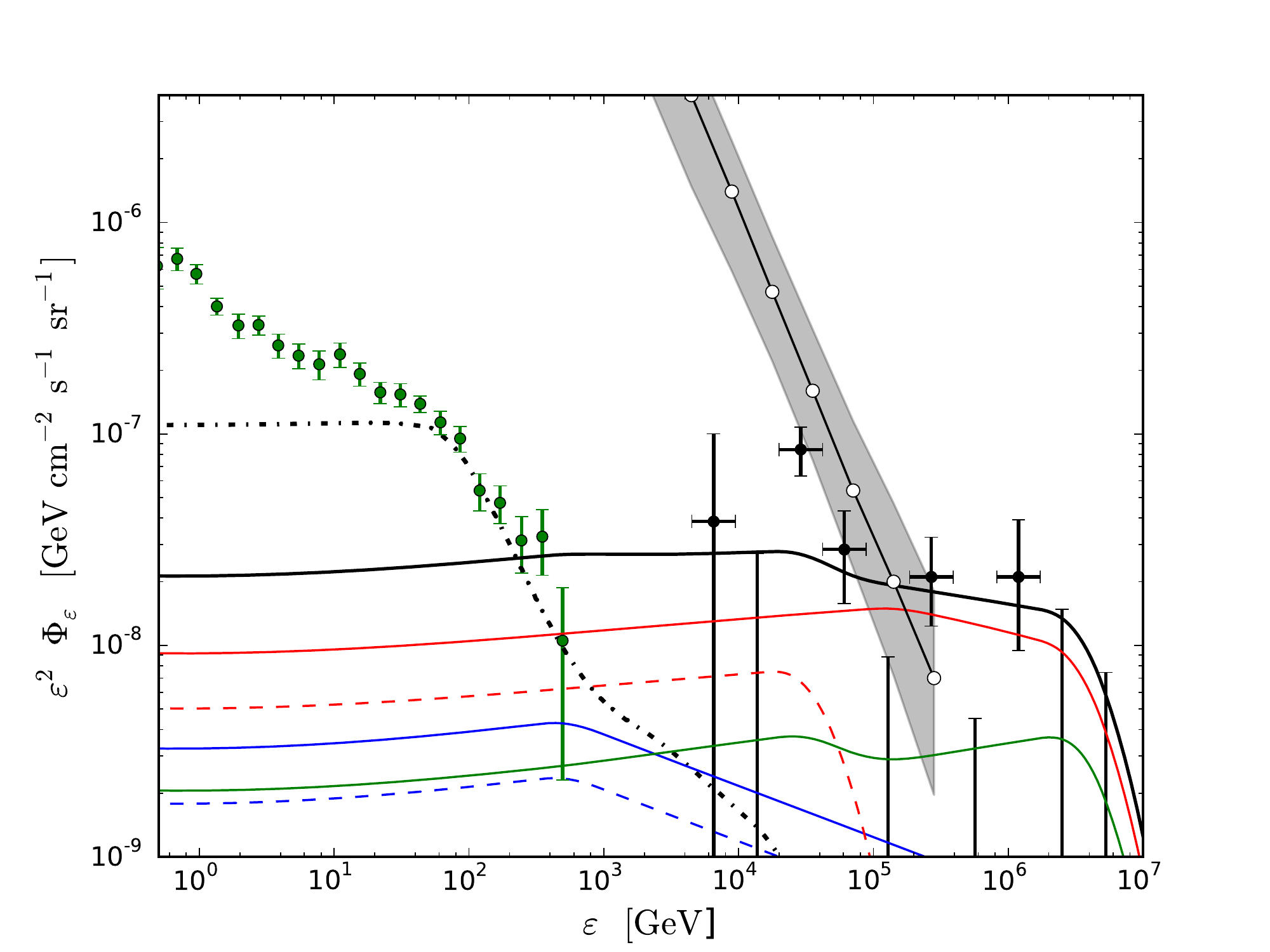}}
\centerline{\includegraphics[width=0.45\textwidth]{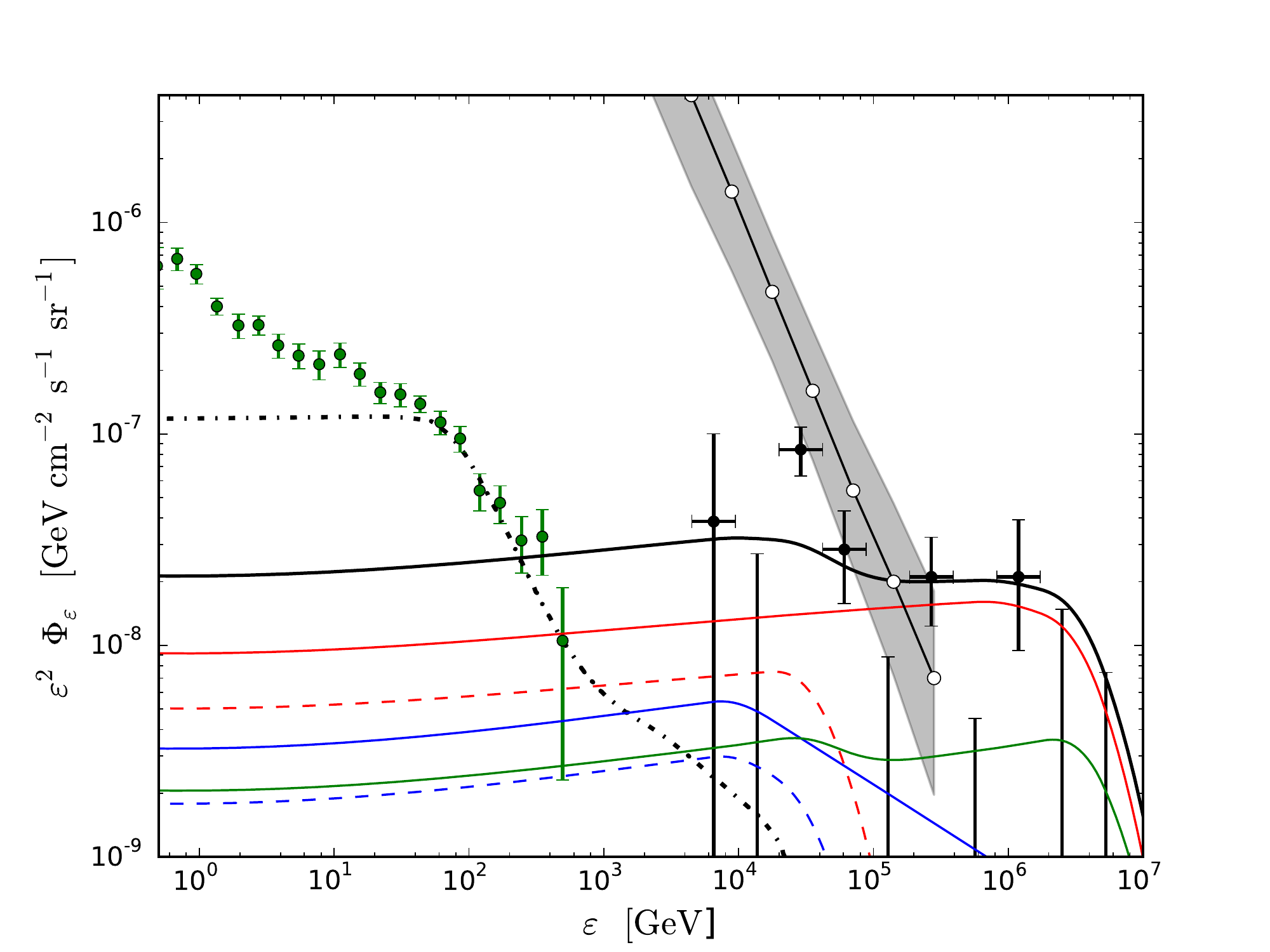}}
\caption{Diffuse flux per flavor of neutrinos (solid black) and $\gamma$-rays (dash-dot) 
from HNe and SNe, for a diffusion coefficient {\bf (top):} $D\propto \vareps_p^{1/3}$,
{\bf (bottom)}: $D\propto \vareps_p^{1/2}$, in both the host galaxy and cluster. 
For both figures HNe and SNe release on average $2.8\times10^{51}$ and $4.8\times10^{49}\;\mathrm{ergs}$ of CR energy respectively, and the proton spectral index is $\Gamma = 2$. The black line with white circles denotes the measured flux of atmospheric neutrinos \citep{IC11atmosphere}. The starburst galaxy scale height, density, and magnetic field strength are 
$H_{sbg} = 300 \pc$, $n_{sbg}=200\cm^{-3}$, and $B_{sbg} = 200\,\mu$G and are represented by red lines. For normal star-forming galaxies $H_{sfg} = 1000\,\pc$, $n_{sfg}=1\cm^{-3}$, and $B_{sfg} = 6\,\mu$G; they are represented by blue lines.
The contribution from HNe are marked with solid lines colored while those from the SNe are dashed. The solid green line denotes the total cluster contribution (i.e. HNe and SNe from both types of galaxies). Green data points correspond to the \textit{Fermi} measurements of the extragalactic diffuse
$\gamma$-ray background \citep{Ackermann+14fermigam}. Black points correspond to the IceCube measurements
of astrophysical neutrinos \citep{IC3+15tevnu} , note that two of the low energy data points are within the gray lines of the error bars of the atmospheric flux. 
}
\label{fig:avg-p2}
\end{figure}
\begin{figure}[h!]
\centerline{\includegraphics[width=0.45\textwidth]{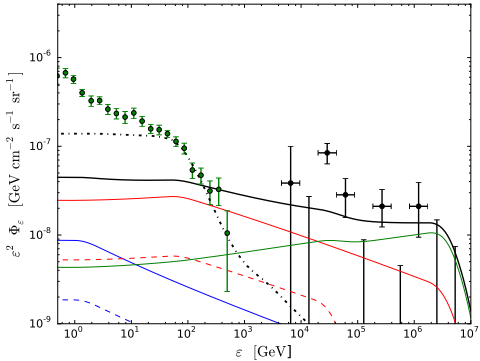}}
\caption{Diffuse flux per flavor of neutrinos (solid black) and $\gamma$-rays (dash-dot) 
from HNe and SNe which release on average $7.5\times10^{51}$ and $5\times10^{49}\;\mathrm{ergs}$ of CR energy respectively, with a phenomenologically motivated diffusion coefficient based on observations of CR diffusion in the Milky Way galaxy (see text for details). 
In this case, cluster contributions are dominant at high energies.}
\label{fig:cluster}
\end{figure}

For our cluster/group parameters and the average galaxy parameters taken in \S \ref{sec:prop},
the diffuse neutrino flux per flavor for a $D\propto \vareps_p^{1/3}$ diffusion coefficient 
is shown in Fig. \ref{fig:avg-p2} (top panel). Here the contributions of the $pp$ interactions 
in the galaxies are indicated both for the supernova and hypernova components.  
In the same figure, the resulting diffuse $\gamma$-ray flux is also shown, resulting from 
the corresponding $\pi^0$ decays and subsequent pair cascades in the intergalactic medium, 
which are  discussed in \S\ref{sec:grcasc}.
A similar calculation for a $D\propto \vareps_p^{1/2}$ Komolgoroff type diffusion coefficient 
is shown in the bottom panel of Fig. \ref{fig:avg-p2}.

The situation depends strongly on the diffusion coefficients of both galaxies and clusters, which are uncertain 
especially at high energies; and, due in large part to uncertainties in the magnetic coherence length.  For example, the diffusion coefficient for normal galaxies used in Fig.~1 
is 10 times lower than the value obtained for our Milky Way. While this discrepancy is alleviated 
by inhomogeneities, the diffuse Galactic emission suggests that the CR spectral break is much lower 
since the observed $\gamma$-ray spectrum is already steep at GeV energies \citep{2012ApJ...750....3A}. 
In Fig.~2, we conservatively use the diffusion coefficient suggested for our Milky Way by \cite{2012JCAP...01..011B} for normal star-forming galaxies. 
We then use the scaling relation $D\propto r_L(\vareps_p)^\alpha$ (see \S \ref{sec:prop}) to determine the diffusion coefficient in starbursts.   Since the break energy is sensitive to the diffusion coefficient, 
one sees that the diffuse neutrino background cannot be explained by star-forming galaxies in this 
case,  even with an optimistically high fraction of HN kinetic energy converted to CRs (i.e. $7.5\times10^{51}\;\mathrm{erg}$). At high energies the galaxy contribution may not be appreciable.
At and below $\vareps_{p\ast,g} \simeq 1.11\times 10^{18}\ell_{g,20}B_{g,-3.7}\eV$, 
however, the galactic contribution becomes considerable, $\tau_{pp,g}(\vareps_{p\ast,g})\simeq 0.33$,
overcoming the cluster contribution at the same energy. For this combination of parameters the cluster and group contribution should be 
dominant, and it is possible to explain the hard spectrum of the diffuse neutrino background. Note that the parameters used for the cluster/group contribution to the diffuse neutrino flux are optimistic, and massive clusters are only a fraction of the 
total cluster population.

\begin{figure}[t]
\centerline{\includegraphics[width=0.45\textwidth]{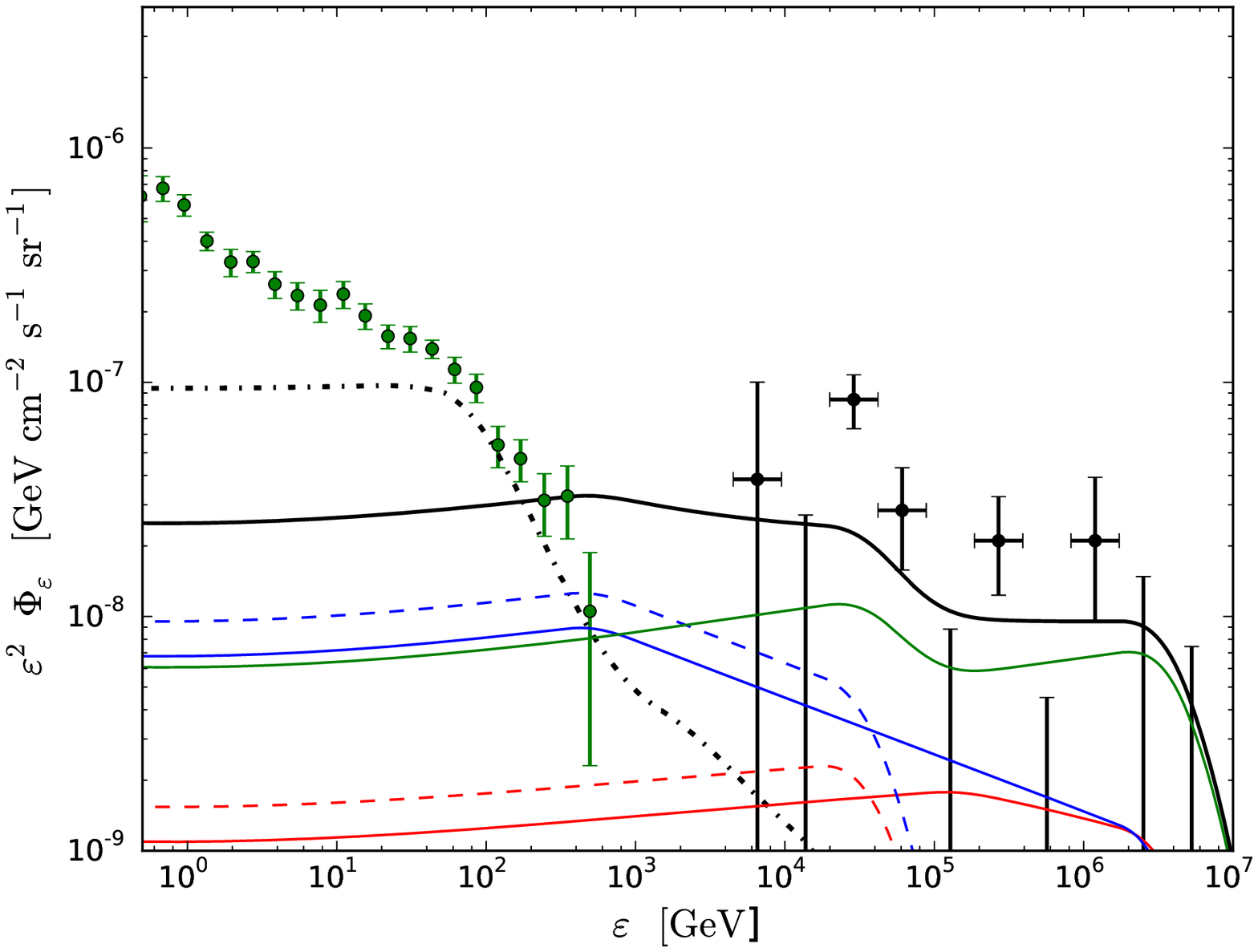}}
\centerline{\includegraphics[width=0.45\textwidth]{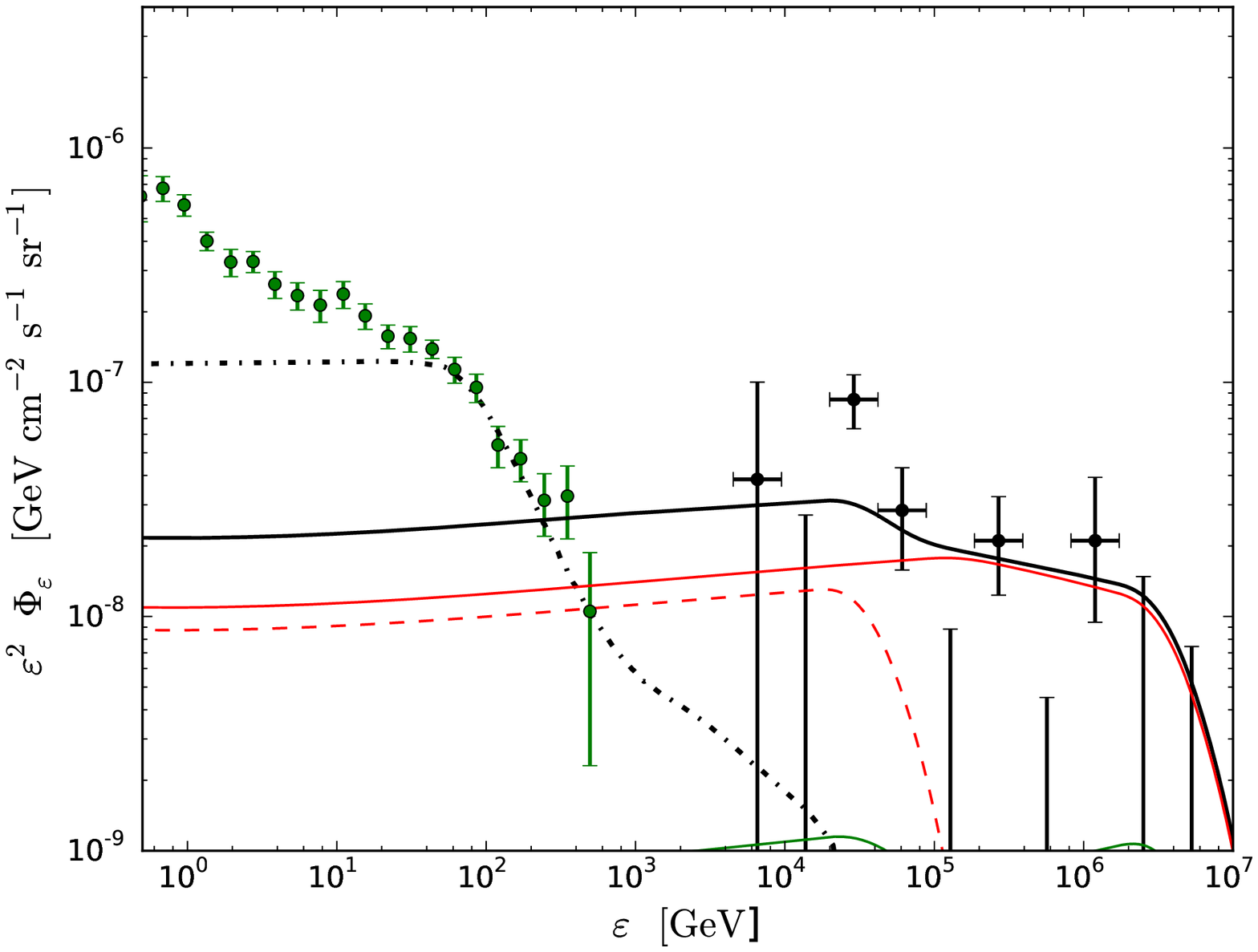}}
\caption{Same as Fig. \ref{fig:avg-p2}, with $D\propto \vareps_p^{1/3}$ but 
{\bf top:}  $\xi_{sbg} = 0.01$ with $E_{cr,hn} = 5\times10^{51}\;\mathrm{erg}$, and $E_{cr,sn} = 2.2\times10^{50}\;\mathrm{erg}$
{\bf bottom:} $\xi_{sbg} = 0.5$ with $E_{cr,hn} = 10^{51}\;\mathrm{erg}$, and $E_{cr,sn} = 2.5\times10^{49}\;\mathrm{erg}$}
\label{fig:nsf_sbg-p2}
\end{figure}

Returning to the parameters used in Fig. \ref{fig:avg-p2}, the flux resulting from average host galaxies with a smaller (top) and larger (bottom) fraction of CRs produced in starburst galaxies is shown in Fig. \ref{fig:nsf_sbg-p2}. Here the fraction of HN/SN CR energy was adjusted ad-hoc in order to fit the observed neutrino flux with $E_{cr,hn} = 5\times10^{51}\;\mathrm{erg}$ and $E_{cr,sn} = 2.2\times10^{50}\;\mathrm{erg}$ for $\xi_{sbg} = 0.01$, and $E_{cr,hn} = 10^{51}\;\mathrm{erg}$ and $E_{cr,sn} = 2.5\times10^{49}\;\mathrm{erg}$ for $\xi_{sbg} = 0.5$ respectively. The diffusion coefficient was taken to be $D\propto \vareps_p^{1/3}$ while leaving the remaining parameters unchanged.

\begin{figure}[t]
\centerline{\includegraphics[width=0.45\textwidth]{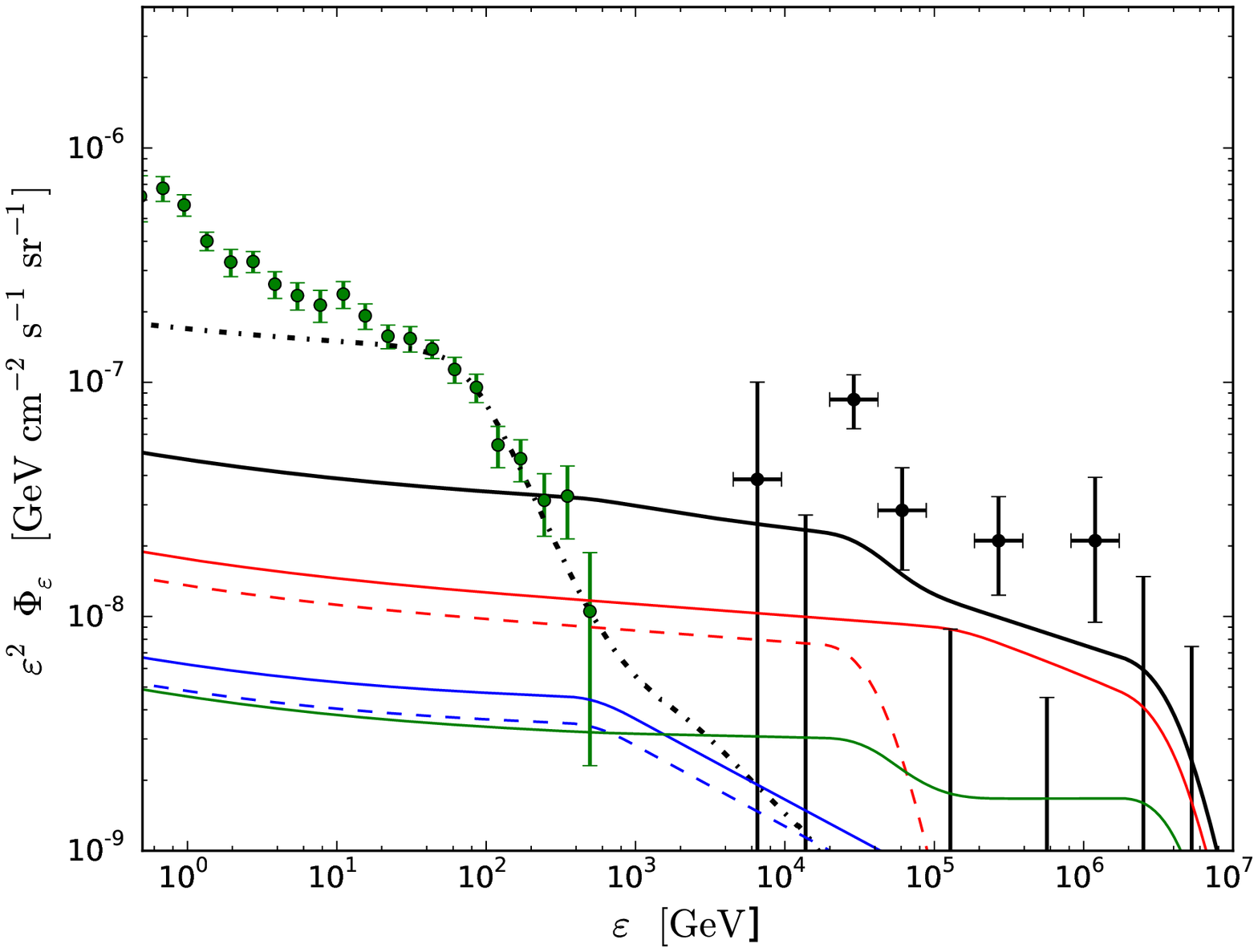}}
\centerline{\includegraphics[width=0.45\textwidth]{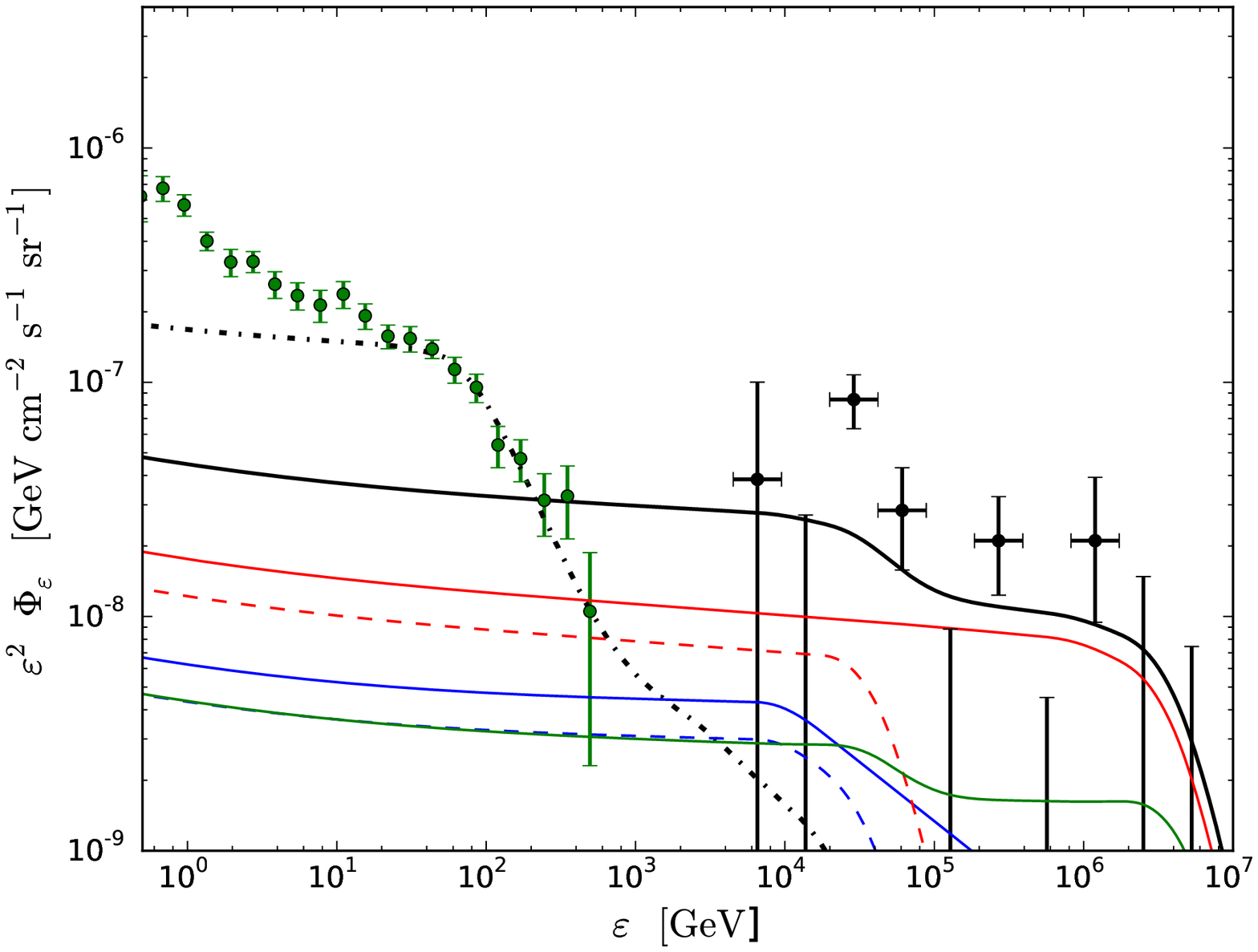}}
\caption{Same as Fig. \ref{fig:avg-p2}, but for a proton index $\Gamma=2.1$, $E_{cr,hn} = 3.5\times10^{51}\;\mathrm{erg}$, and $E_{cr,sn} = 
10^{50}\;\mathrm{erg}$.
{\bf top:} $D\propto \vareps_p^{1/3}$ ,
{\bf bottom:} $D\propto \vareps_p^{1/2}$.
}
\label{fig:soft-p2}
\end{figure}

The Fig. \ref{fig:avg-p2} and \ref{fig:nsf_sbg-p2} were calculated for ``typical" star-forming
galaxies with parameters as given above, and for a proton injection spectrum $\Gamma=2$.  We consider next
the SFG and SBG contributions using the same parameters, but with a proton injection
index $\Gamma=2.1$, results are shown in Fig. \ref{fig:soft-p2}.

The effect of $p\gamma$ interactions in the galactic and intracluster medium is sub-dominant 
relative to the $pp$ collisions in the relevant energy range, although it becomes dominant at 
very high energies \citep{2009ApJ...707..370K}.

%
%

%

\section{Gamma-Ray Cascades} 
\label{sec:grcasc}

The same $pp$ interactions which produce neutrinos also produce high-energy $\gamma$-rays with typical energy $\epsilon_\gamma \sim  \vareps_p/10$. Note that because of this connection, their resulting flux can be related by $\epsilon_\gamma^2\Phi_{\epsilon_\gamma} = \left. 2^{\Gamma-1}\,\varepsilon^2_\nu\Phi_{\varepsilon_\nu}\right|_{\varepsilon_\nu = 0.5\,\epsilon_\gamma}$.   
When $\gamma$-rays with energy $\epsilon'_\gamma \gtrsim 100$ GeV are injected into intergalactic space sufficiently far from Earth (i.e. $\sim$ few Mpc), they undergo $\gamma\gamma$ interactions with extragalactic background light (EBL) photons producing electron/position pairs. The pairs scatter additional EBL photons via the inverse Compton mechanism generating an electromagnetic cascade. The resulting cascaded $\gamma$-ray spectrum takes a universal form, \citep[e.g.,][]{Berezinskii+75cascades,Murase+12nugamcasc}:
\begin{equation}
\label{eq:cascade}
\epsilon_\gamma\frac{dN_\gamma}{d\epsilon_\gamma}  \propto G_{\epsilon_\gamma} = \left\{
\quad\begin{array}{l l}
\left(\frac{\epsilon_\gamma}{\epsilon^{br}_\gamma}\right)^{-1/2} &\; \epsilon_\gamma \leq \epsilon_\gamma^{br}\\
&\\
\left(\frac{\epsilon_\gamma}{\epsilon^{br}_\gamma}\right)^{1-s} &\;  \epsilon_\gamma^{br} < \epsilon_\gamma \leq \epsilon_\gamma^{cut}
\end{array}
\right.
\end{equation}
The characteristic energies \Ecut\, and \Ebreak\, given above are defined by $1 =\tau_{\gamma\gamma}\left[\epsilon^{cut}_\gamma,z\right]$ and \Ebreak = $0.0085\,(1+z)^2\, \left(\frac{\epsilon_\gamma^{cut}}{0.1\;\mathrm{TeV}}\right)^2$ respectively and the high-energy spectral index is generally taken to be $s\sim 2$. Here $\tau_{\gamma\gamma}$ is the optical depth for a high energy photon traveling through intergalactic space, the values for which are from model C of \cite{Finke+09eblmodel}. 

There is also an attenuated component to the observed $\gamma$-ray flux from photons with energy $\epsilon_\gamma \lesssim $ \Ecut which can be calculated similarly to eq. (\ref{eq:fluxnugen})
\bea
&\epsilon_\gamma^2\Phi^{unatt}_\gamma =  & \frac{c}{4\pi} \int dz \left|\frac{dt}{dz}\right| 
 e^{-\tau_{\gamma\gamma}[(1+z)\,\epsilon_\gamma,z]}\; \times\nonumber\\
 & \frac{1}{(1+z)^4} \Bigg[ \frac{2^{\Gamma-2}}{3}\,  & \left. \sum_i f_{i,pp}\, \left(\vareps_p\,Q_{\vareps_p}\right)_{phys} \right]_{\varepsilon'_{cr}=10 (1+z)\epsilon_\gamma} 
\label{eq:fluxggen}
\ena
which combined with Eq. (\ref{eq:cascade}) can be compared with \textit{Fermi}-LAT measurements of 
the extragalactic diffuse $\gamma$-ray background \citep{Ackermann+14fermigam}. Figures 
\ref{fig:avg-p2} through \ref{fig:soft-p2} show our 
calculated diffuse flux of neutrinos and $\gamma$-rays along with data from IceCube and \textit{Fermi}.

\section{Discussion and Summary}
\label{sec:disc}
In this work, we discuss the starburst scenario in light of the new 10 TeV neutrino data. 
Although there are large systematic uncertainties involved in removing the atmospheric muon background at such low energies \citep{IC3+14tevnu}, 
it may be challenging to explain the diffuse neutrino flux in the whole energy range with a single power-law 
component with $\Gamma \sim 2$. Adding the SNe contribution enables us to explain the low-energy data, but we 
find that constraints from the diffuse $\gamma$-ray background are quite stringent \citep{Murase+13pev}. If the CR energy 
input by SNe is a factor of two larger than that by HNe, the diffuse $\gamma$-ray background 
is violated. Additional constraints could be placed on the cluster contribution by considering the concomitant radio emission, although in the strong evolution case these limits  are weaker than the ones imposed by the accretion shock scenario \citep{2014arXiv1410.8697Z}. From the $\gamma$-ray limits, we conclude that if the diffuse neutrino background in the PeV range originates 
mainly from HNe (and their host galaxies), the HN contribution should be larger than or at least 
comparable to the SN contribution.    
However, interestingly, in cases where the cluster/group contribution is mainly responsible for the diffuse neutrino 
flux, it is still possible for the SN contribution to overwhelm the HN contribution (e.g. the top panel of Fig. \ref{fig:nsf_sbg-p2}). 

The strong case scenario, where the $\sim10$ TeV neutrino data are explained by CR reservoirs, has an interesting feature 
that can be tested soon.  As proposed by \citet{Murase+13pev}, CR reservoirs can give a common explanation 
for both the diffuse neutrino and $\gamma$-ray backgrounds.  In general, the contributions from 
starbursts and other sources to the neutrino flux above 100 TeV  result in subdominant contributions to the diffuse $\gamma$-ray background. However, as we show above any source which contributes significantly to the 10 TeV diffuse neutrino flux in the $pp$ scenario must also account for almost all of the diffuse $\gamma$-ray background.  
It is commonly believed that the diffuse $\gamma$-ray background is dominated by unresolved blazars 
\citep{2014arXiv1412.3886I,2014JCAP...11..021D},  implying a comparatively smaller starburst contribution.  
Although there are still significant uncertainties in the modeling of both blazar \citep[e.g.,][]{2014ApJ...780...73A,2012ApJ...751..108A} and starburst contributions \citep[e.g.,][]{2012ApJ...755..164A,Tamborra+14igbnu}, our results imply that the strong case scenario can 
be tested by an improved understanding or characterization of the diffuse $\gamma$-ray background. 

If, for example, it is proven that blazars are responsible for $\gtrsim50$\% of the observed diffuse $\gamma$-ray 
background, the starburst contribution to the diffuse neutrino background at low energies should be small, 
especially if the CR energy input from SNe is comparable to or larger than that from HNe. Specifically, tighter constraints on the the unresolved blazar fraction of the diffuse $\gamma$-ray background measured by \textit{Fermi} and possibly the High Altitude Water Cherenkov (HAWC) observatory \citep{benzvi2015HAWCresults}, as well as the low energy ($\varepsilon_\nu \lesssim 100\;\mathrm{TeV}$) spectral shape of the astrophysical neutrino flux as analyzed by \cite{IC3+14tevnu}, will impose limits to the corresponding fluxes from galaxy clusters/groups.
If there is little room for the CR reservoirs, other sources need to be responsible for the low-energy neutrino component. For instance, there might be a significant contribution from Galactic sources. Although the Galactic diffuse emission by CRs propagating in our Milky Way cannot provide the main contribution \citep{Ahlers+14pevnugal,2014MNRAS.439.3414J}, 
some extended sources such as the Fermi Bubbles \citep{Ahlers+14pevnugal,2014PhRvD..90b3016L} or 
nearby HN remnants \citep{Fox+13pev,Ahlers+14pevnugal} could be viable. 
Alternatively, the diffuse neutrino background might be produced mainly by hidden neutrino sources 
via $p\gamma$ processes \citep{2013PhRvL.111l1102M,2014arXiv1410.2600K,2014arXiv1411.3588K}.  
The advantage of the strong case considered here is that it can be tested by multimessenger approaches.    
It has been commonly believed that Galactic CRs come from SN remnants. If the diffuse neutrino 
background is dominated by star-forming galaxies, our results imply that even Galactic CRs may include
significant contributions from past HN remnants.   

CR acceleration to energies $\simg {10}^{16}-10^{17}\eV$ has also been proposed in other accelerators, 
such as shocks in AGN jets, \citep[e.g.,][]{Hillas84rev,Biermann+87agncr,2014JHEAp...3...29D}, 
AGN winds \citep{Murase+14nuagn,Tamborra+14igbnu}, and AGN cores 
\citep{1991PhRvL..66.2697S,2004PhRvD..70l3001A,2014arXiv1411.3588K}. 
While in such cases neutrinos can come from AGNs themselves, CRs escaping from AGNs can also produce neutrinos 
in intergalactic space, which may give a significant contribution to the diffuse neutrino 
background \citep{Murase+13pev}.  Other possibilities are accretion shocks onto clusters 
of galaxies \citep{Murase+08clusternu,2014arXiv1412.5678D}, galaxy mergers in clusters 
\citep{Kashiyama+14pevmerg}, and $\gamma$-ray bursts including trans-relativistic SNe or low-luminosity $\gamma$-ray bursts 
\citep{2008PhRvD..78b3005M,2007PhRvD..76h3009W}.  
In principle, the discussion of the cluster/group propagation 
effects discussed above also applies to any intracluster sources. 
The main difference between these other sources and SNe/HNe (or sources inside 
galaxies in general) is that the CRs accelerated in the former do not undergo $pp$ interactions 
in the galactic gas, but only in the intracluster gas; whereas CRs from HNe and galactic sources 
undergo $pp$ interactions in both the host galaxy and the cluster/group. This disparity may be 
relevant at sub-PeV and TeV energies, where the spectral shape of the neutrino flux can provide clues to 
the source. In this energy range we expect the advective escape and 
Hubble times to dominate the galactic and cluster diffusion times respectively, and at different 
critical energies (i.e. $\vareps_{w}$ and $\vareps_{p,cl}$ as in \S \ref{sec:prop}).  
Therefore, for galactic sources (that are extragalactic), a soft spectrum is typically expected at energies below the maximum
acceleration energy, with $\vareps_\nu^2\Phi_\nu \propto \vareps_\nu^{-\alpha}$ (for a diffusion
time $\propto \vareps_\nu^{-\alpha}$), with a leveling off of the slope to $\vareps_\nu^2\Phi_\nu 
\sim$constant below about $\vareps_{\nu,g} \sim 130\,\left(2/(1+z)\right)$ TeV (assuming, e.g., 
a galactic magnetic field strength of $200\mu$G and diffusion exponent $\alpha = 1/2$). 
Sources which release their CRs directly into the intracluster medium on the other hand, are expected 
to produce relatively flat neutrino spectra below a break around $\vareps_{\nu,cl} \sim \mathrm{few}\,
\left(2/(1+z)\right)$ PeV, steepening above that. Such a spectral softening of the cluster contribution is not seen for the HNe model considered here (e.g. in Fig. \ref{fig:cluster}) because the break energy due to diffusion is lower than the cutoff energy imposed by maximally accelerated CRs.
While both $\vareps_{\nu,g}$ and $\vareps_{\nu,cl}$ are subject to large uncertainties in parameters, the presence of their corresponding features could be suggestive of sources which are embedded in star-forming or starburst galaxies.

As shown by \cite{Murase+13pev}, in the $pp$ scenario, the neutrino spectrum cannot be 
softer than about $\vareps_\nu^2\Phi_\nu \propto \vareps_\nu^{-0.2}$ at low energies for the 
corresponding $\gamma$-ray component not to violate the \textit{Fermi} 
measurements of the diffuse $\gamma$-ray background \citep{Ackermann+14fermigam}.    
At the same time, a flat spectrum at moderate to high energies creates 
tension with the non-detection of neutrinos with energies near the Glashow resonance at $\vareps_\nu 
\sim 6$ PeV, which necessitates a neutrino spectral shape near that energy steeper than 
$\vareps_\nu^2\Phi_\nu \propto \vareps_\nu^{-0.3}$ \citep{2013PhRvD..88d3009L}. 
Such a spectral break can occur if acceleration 
stops or there is a transition in the diffusive escape time (i.e. $D(\vareps_{cr})\propto 
\vareps_{cr}^\alpha\to D(\vareps_{cr})\propto \vareps_{cr}^2$) around 
$\vareps_{cr}\sim 240\,\left((1+z)/2\right)$ PeV  \citep[e.g.,][]{Murase+13pev,Anchordoqui+14pevnu}.
As can be seen in Fig. \ref{fig:avg-p2} and \ref{fig:nsf_sbg-p2}, 
the model presented here can also resolve this tension.  The neutrino spectrum is flat at 
low energies $\vareps_\nu \lesssim 130\TeV$ and softens to $\vareps_\nu^2\Phi_\nu \propto 
\vareps_\nu^{-\alpha}$ slightly before the Glashow resonance, while the corresponding diffuse 
$\gamma$-ray spectrum is below the \textit{Fermi} measured flux.

We have shown in \S \ref{sec:difnu} that the high-energy diffuse neutrino flux could potentially 
be explained by HNe, predominantly those located in dense starburst galaxies (e.g. the red 
solid curve in the bottom panel of Fig. \ref{fig:avg-p2}) especially for 
$\vareps_\nu \gtrsim 50$ TeV. For neutrinos with this energy and below, the SNe in both starburst 
and normal star-forming galaxies contribute significantly to the diffuse flux and produce a ``bump'' in 
the spectrum\footnote{As we were preparing to submit our calculation, a similar qualitative
conclusion was posted by \cite{Chakraborty+15tevpevnu}, who however did not consider the diffuse 
$\gamma$-ray constraints from {\it Fermi} observations that are very important.}. Reasonable fits by eye are
obtained for the diffuse neutrino flux measured by IceCube including the latest TeV data by using reasonable parameters for 
the sources as well as the diffusion properties in hosting structures. 
Such a flux also approximates but does not violate the diffuse $\gamma$-ray background measured by \textit{Fermi}. 
This does not mean that SNe and HNe are necessarily the only sources contributing to the neutrino and 
$\gamma$-ray diffuse backgrounds. It supports, however, the case for CR reservoirs such as clusters
and groups being promising sources which could contributes at least a significant fraction of these 
backgrounds, without violating both CR \citep{2014PhRvD..90l3012Y} and $\gamma$-ray constraints.

\acknowledgments{This work was partially supported by NASA NNX13AH50G. 
We are grateful to J. van Santen for providing the IceCube data points.

\bibliographystyle{apj}
\bibliography{ms_accepted.bbl}

\end{document}